\documentclass[preprint,12pt]{aastex}               % 1 column AAS preprint
\usepackage{epsfig}

\def\beq{\begin{equation}}
\def\eeq{\end{equation}}

\begin{document}

%\shorttitle {Magnetic Reconnection in Astrophysical Systems}
%\shortauthors{D.~A.~Uzdensky}

\title{Magnetic Reconnection in Astrophysical Systems}
\author{Dmitri A.\ Uzdensky\altaffilmark{1}}
\altaffiltext{1}{Department of Astrophysical Sciences, Princeton University, 
Peyton Hall, Princeton, NJ 08544 --- Center for Magnetic Self-Organization
(CMSO); {\tt uzdensky@astro.princeton.edu}.}

%\date{\today}
\date{July 28, 2006}

\begin{abstract}
The main subject of my talk is the question: in what kind of 
astrophysical systems magnetic reconnection is interesting and/or 
important? To address this question, I first put forward three 
general criteria for selecting the relevant astrophysical environments. 
Namely, reconnection should be: fast; energetically important; and 
observable. From this, I deduce that the gas density should be low, 
so that the plasma is: collisionless; force-free; and optically thin. 
Thus, for example, the requirement that reconnection is fast implies 
that Petschek's reconnection mechanism must be operating, which is
possible, apparently, only in the collisionless regime. Next, I 
argue that the force-free condition implies that the magnetic field 
be produced in, and anchored by, a nearby dense massive object, e.g., 
a star or a disk, strongly stratified by gravity. I then stress the 
importance of field-line opening (e.g., by differential rotation) as 
a means to form a reconnecting current sheet. Correspondingly, I 
suggest the Y-point helmet streamer as a generic prototypical magnetic 
configuration relevant to large-scale reconnection in astrophysics.
Finally, I discuss several specific astrophysical systems where the 
above criteria are met: stellar coronae, magnetically-interacting 
star--disk systems, and magnetized coronae above turbulent accretion 
disks. In the Appendix I apply the ideas put forward in this talk to
the solar coronal heating problem.
\end{abstract}

%*****************************************************************

\section {Introduction}
\label{sec-intro}

%------------------------------------------------------------------

\subsection{Warnings and Disclaimers}
\label{subsec-disclaimers}

This writing is loosely based on an invited talk the author gave at 
the Harry S.~Petschek Memorial Symposium at the University of Maryland
(College Park, Maryland, March 21--23, 2006).

What I am going to talk about today is not going to be universally 
applicable to everything in astrophysics, but I just want to advance 
a certain perspective, a certain line of reasoning that, I think, 
should be considered mainstream. So mainstream in fact, that it may 
appear trivial. But I think it is worth to systematically re-emphasize
some basic things from time to time. 

Thus, I would like to start out by warning the audience that what 
I am going to say may be found either trivial or simply incorrect. 
And rightfully so. Nevertheless, here I go.

From the start, I would like to make one important reservation.
My discussion applies only to individually-discernible, large-scale, 
energetic reconnection events (I shall call them flares), and not 
to a background of numerous little reconnections that may be encountered, 
for example, at the bottom of a turbulent MHD cascade. Such small-scale 
reconnections (e.g., nano-flares) may be even ultimately responsible for 
the bulk of energy dissipation in MHD turbulence. However, I will not 
discuss them in this talk.

My final disclaimer is about completeness of my reference list.
My chioce of toices and astrophysical examples is rather arbitrary
and I don't claim it to be comprehensive. Similarly, my reference 
list is not complete and does not pretend to be complete.

%------------------------------------------------------------------

\subsection{Talk Outline}
\label{subsec-outline}

The subject of this session and the subject of my talk is 
{\it Reconnection in Astrophysical Systems}. So, when I started 
preparing this talk, I first had to ask myself: well, indeed, 
{\it in what astrophysical systems does it make sense to talk 
about magnetic reconnection?}
So, my talk is going to be devoted entirely to trying to answer 
this question.

In the first part of my talk I will put forward the following
three general criteria that I think one should use when selecting
reconnecting astrophysical systems. Reconnection should be:
{\bf
\begin{enumerate}
\item{fast}
\item{energetically important}
\item{observable}
\end{enumerate}
}

I will then go briefly through each of these conditions and discuss 
what each of them means for the physical environment in question. I 
will argue that these criteria can be translated into the following 
three physical requirements:
{\bf
\begin{enumerate}
\item{collisionless}
\item{force-free}
\item{optically-thin}
\end{enumerate}
}

In the second part of my talk I will apply these criteria to determine 
logically in what types of astrophysical systems reconnection plays a 
role and should be studied. I will first argue that, generically, the 
force-free requirement implies that one deals with a magnetosphere or 
a corona of a nearby colder and denser gravitating object (e.g., a star 
or an accretion disk), with gravity being responsible for gas stratification. 
The magnetic field originates in this dense plasma. Then I will discuss
field-line opening (driven by sheared footpoint motion, a wind, or by 
relativistic effects) as a common and natural way to create a large-scale 
current sheet in the coronal plasma, which is necessary for reconnection. 
Correspondingly, I will advocate a {\it Y-point Helmet Streamer} as one 
of the most important and generic magnetic structures relevant to 
astrophysical reconnection.

In the last part of my talk, I will discuss several specific astrophysical 
examples where reconnection is believed to be playing an important role.

Finally, in the Appendix I describe solar coronal heating as 
a self-regulated process keeping the coronal plasma marginally 
collisionless.

%****************************************************************

\section{Fast Reconnection}
\label{sec-fast}

First, what is {\it fast reconnection}? Usually in the magnetic reconnection 
literature, a reconnection mechanism is called ``fast'' if the reconnection 
rate is independent (or scales only logarithmically with) the classical
resistivity. In my talk, I will use a somewhat broader definition. I will 
call a reconnection process fast when the reconnection rate (defined as 
the ratio of reconnection velocity to the Alfv\'en speed) is independent 
of (or depends only relatively weakly on) the global size~$L$ of the system. 
In other words, in this definition reconnection is fast when its dimensionless 
rate is determined (almost) entirely by the local physical parameters near 
the center of the reconnection layer. In this sense, for example, the 
classical Sweet--Parker reconnection is slow, because its rate scales 
as~$L^{1/2}$, wheres the maximum Petschek reconnection rate is (almost) 
fast, since it scales only logarithmically with~$L$.

%------------------------------------------------------------------

\subsection{Main Mechanism of Fast Reconnection in Astrophysics}
\label{subsec-handwaves}

In Astrophysics, {\it reconnection} is a magic word
(a ``tooth fairy'', as it would be called in Peyton Hall),%
\footnote
{Another well-known astrophysical ``tooth fairy'' is {\it magnetic field} 
itself.}
in the sense that it has become customary to invoke reconnection when 
it is needed to solve one's problems and to assume that it always works 
when called upon.

It has to be noted that the main reconnection mechanism in 
Astrophysics is NOT Petschek reconnection, nor is it Hall 
reconnection, nor anomalous-resistivity reconnection. No, 
the most important reconnection mechanism in Astrophysics 
invokes waves, a certain type of waves, in fact. Called 
{\it handwaves}%
\footnote
{I acknowledge first hearing a satyrical mentioning of handwaves as a playful 
``real'' physical mechanism from Henk Spruit (2005, private communication).}
(See Fig.~1). The mechanism works like this: 
{\it Well, we know that fast reconnection happens in 
the Solar corona, and in the Earth magnetosphere. So 
it should also happen in OUR astrophysical system.}

Following this well-established and respected astrophysical tradition, 
I will also make extensive use of hand-waving arguments throughout my 
talk :)

\begin{figure} [h]
\centerline{\psfig{file=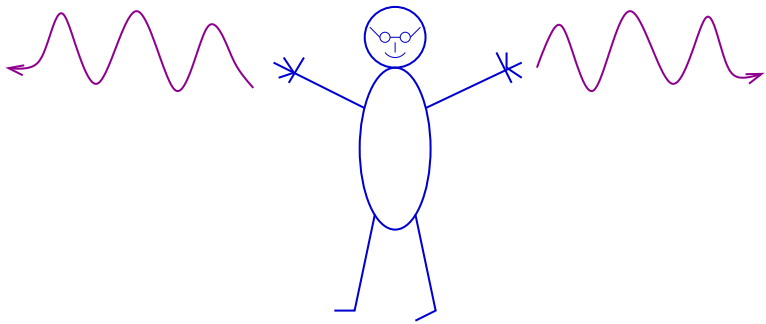,width=4 in}}
\figcaption{Main Reconnection Mechanism in Astrophysics.
\label{fig-handwaves}}
\end{figure}

%--------------------------------------------------------

\subsection{Fast Reconnection: Petschek's Legacy}
\label{subsec-Petschek}

Those who are not satisfied with the mechanism described in the 
previous subsection, have to rely on actual thought. An excellent 
example of a brilliant thinker, who contributed a lot to our 
understanding of reconnection, is Harry Petschek. Why are Petschek's 
ideas on reconnection important? People now associate fast reconnection 
with Petschek's (1964) reconnection mechanism.%
\footnote
{By the way, I will assume the audience to be familiar with 
both the Sweet-Parker and Petschek models of reconnection.}
A great non-trivial idea due to Petschek is that the main bottleneck 
stifling the reconnection process in the classical Sweet--Parker
(Sweet~1958; Parker~1957) model can be circumvented if one can set 
up a special magnetic field and flow configuration --- the Petschek 
configuration. The bottleneck arises because of the necessity to have 
a reconnection layer that is simultaneously thin enough for the resistivity 
to be important and thick enough for the plasma to be able to flow out 
of the layer. Petschek's key idea was that the thin current layer and 
the thick outflow channel do not have to be the same if the reconnection 
region is not a simple rectangular box, as it was in the Sweet--Parker 
theory, but has a somewhat more complicated structure, with four standing 
slow shocks attached to a central diffusion region. As a result, there is 
an additional geometric factor that can lead to faster reconnection. 

This geometrical enhancement is especially important in astrophysics, for 
the following reason. What distinguishes astrophysical and space systems 
from Earth-bound laboratory experiments is a huge contrast in length-scales.
Astronomical systems are astronomically large: the system size~$L$ is 
usually much greater than the microscopic physical scales, e.g., the ion 
gyro-radius~$\rho_i$, the ion collisionless skin-depth~$d_i$, and the 
Sweet--Parker reconnection layer thickness~$\delta_{\rm SP}$. Hence, the 
reconnection rate problem is especially severe in astrophysical systems. 
Hence, we need a clever idea. For example, the idea of geometric 
enhancement due to Harry Petschek.

What this means is that, unless one has a special mechanism like this, 
no microphysics (e.g., the Hall effect or anomalous resistivity) can 
give reconnection rates that are rapid enough to be of interest to 
observations. For example, when the reconnection layer's thickness 
becomes comparable with the collisionless ion skin depth~$d_i$, the 
layer enters the Hall regime. Then, in a simple Sweet--Parker-like analysis,
the mass conservation condition would result in a reconnection velocity that 
is by a factor of~$d_i/L$ smaller than the Alfv\'en speed~$V_A$. Since $d_i$ 
is usually much smaller than~$L$, this rate would be much too slow to be of 
practical interest.

%--------------------------------------------------------

\subsection{Fast Reconnection Means Collisionless Reconnection}
\label{subsec-collisionless}

Thus, we see that Petschek's mechanism, or a variation thereof, 
is absolutely indispensable for astrophysical reconnection.
Unfortunately, however, several numerical and analytical studies 
(e.g., Biskamp~1986; Scholer~1989; Uzdensky \& Kulsrud~2000; 
Erkaev~et~al. 2001; Kulsrud~2001; Malyshkin et al.~2005) 
have shown that in resistive MHD with uniform resistivity (and, 
by inference, with resistivity that is a smooth function of 
plasma parameters, e.g., Spitzer) Petschek's mechanism fails 
and Sweet--Parker scaling applies instead. The same conclusion
was achieved in laboratory studies by the Magnetic Reconnection 
Experiment (MRX), lead by Masaaki Yamada at Princeton Plasma 
Physics Laboratory, in the high-collisionality regime (Ji~et~al.~1998;
Trintchouk~et~al.~2003). 

What all this means is that, whenever classical resistive MHD applies, 
one does not get fast reconnection. This implies that fast reconnection 
can happen {\it only} when the plasma is relatively collisionless so that
resistive MHD doesn't apply. This condition of fast reconnection can be 
formulated roughly% 
\footnote
{If the guide component of the magnetic field is not zero,
this condition may be somewhat different, although similar
in concept. For simplicity, however, in this paper we shall 
consider the fast reconnection condition only as given by 
equation~(\ref{eq-1}).} 
as (e.g., Yamada~et~al. 2006)
\beq
\delta_{\rm SP}\ll d_i \equiv {c\over{\omega_{pi}}} \, .
\label{eq-1}
\eeq

What this condition means is the following. As a reconnection layer 
is forming, its thickness $\delta$ is getting smaller and smaller. 
If condition~(\ref{eq-1}) is not satisfied, then this thinning saturates 
at $\delta=\delta_{\rm SP}$, and reconnection then proceeds in the slow 
Sweet--Parker regime.
However, if condition~(\ref{eq-1}) is satisfied, then various two-fluid 
and/or kinetic effects kick in as soon as~$\delta$ drops down to about~$d_i$ 
or so, well before the collisional resistive effects become important.
Then, the reconnection procesess necessarily involves collisionless,
non-classical-resistive-MHD physics.

%---------------------------------------------------------------------

Thus, in the collisional regime, when classical resistive MHD applies, 
fast Petschek reconnection does not appear to be possible. Does going 
to the collisionless regime help? There is a growing consensus that 
the answer to this question is YES. 
In Space/Solar physics, of course, there has long been a very 
serious evidence for fast collisionless reconnection; it has 
been further significantly strengthened by recent laboratory
measurements in the MRX (Ji~et~al. 1998; Yamada~et~al. 2006). 
These measurements, however, have not been able to elucidate 
the special role of the Petschek mechanism in accelerating 
reconnection. On the other hand, over the past decade or so, 
several theoretical and numerical studies have indicated that 
fast reconnection enhanced by the Petschek mechanism (or a 
variation thereof) does indeed take place in the collisionless 
regime. It appears that there may be two regimes of collisionless 
reconnection. Physically, these two possibilities are very different 
from each other; nevertheless, they both appear to lead to the 
establishment of a Petschek-like configuration, which enhances 
the reconnection rate. The two regimes in question are:

\begin{itemize}
\item{{\it Hall-MHD reconnection}, involving two-fluid effects in 
a {\it laminar} flow configuration (e.g., Shay~et~al. 1998; Birn~et~al.~2001; 
Bhattacharjee~et~al. 2001). [See, however, recent particle simulations by 
Daughton et~al. (2006) and by Fujimoto (2006) that cast doubt on Hall 
reconnection as a possible fast reconnection mechanism.]}  
\item{Spatially-localized {\it anomalous resistivity} due to micro-turbulence; 
this seems to lead to a Petschek configuration with the inner diffusion 
region having a width of the order of the resistivity localization scale
(e.g., Ugai \& Tsuda 1977; Sato \& Hayashi 1979; Scholer~1989; Biskamp \& 
Schwarz 2001; Erkaev~et~al. 2001; Kulsrud~2001; Malyshkin~et~al. 2005).}
\end{itemize}

At present, it is still not clear which one of these two mechanisms 
works in a given physical situation (if at all). Also not known is
whether these two mechanisms can coexist and perhaps even enhance 
each other. Recent experimental evidence from the MRX experiment 
suggests that both regimes do exist in reality and that they may
operate simultaneously in a given system (Yamada~2006, private 
communication).

In any case, it seems that one does get a Petschek-enhanced fast 
reconnection process if the plasma is collisionless [in the sense
of equation~(\ref{eq-3})]. To sum up, in order for astrophysical 
reconnection to be fast, it needs Petschek's mechanism to operate 
and that in turn requires the reconnection layer to be collisionless. 
Thus, for the purposes of this talk, I will put an equal sign between 
collisionless reconnection and fast Petschek's reconnection.

In fact, whenever we observe violent and rapid energetic phenomena
that we interpret as reconnection, it is always in relatively tenuous 
plasmas. Please correct me if this is not so. I would be very interested 
in learning about counter-examples. Is there any evidence for fast 
large-scale reconnection events in collisional astrophysical environments?

%---------------------------------------------------------------------

\subsection{Fast Reconnection: Range of Densities in Astrophysics}
\label{subsec-densities}

Note that the statement of collisionality is scale-dependent. This 
is because~$d_i$ is a microscopic scale, independent of~$L$, whereas 
$\delta_{\rm SP}\sim\sqrt{L}$. In astrophysics $L\gg d_i$ is very 
large, and so one might expect $d_i\ll \delta_{\rm SP}$ for large 
enough systems. However, in practice, this doesn't always have to 
be so. Indeed, notice that $\delta_{\rm SP}\sim 1/\sqrt{S} \sim 
V_A^{-1/2} \sim \rho^{-1/4}$, whereas $d_i\sim\rho^{-1/2}$. Thus, 
$$
{{\delta_{\rm SP}}\over{d_i}} \sim \rho^{1/4} \, .
$$
so it depends on density. This dependence is weak but, in astrophysics, 
one has to deal not only with a huge dynamic range in~$L$ but also with 
an enormous dynamic range in~$\rho$. For example,
\begin{itemize}
\item{$\rho\sim 10^{14}\,{\rm g/cm}^3$ inside a neutron star;} 
\item{$\rho\sim 1\,{\rm g/cm}^3$ average solar density;}
\item{$n_e\sim 10^{10}\,{\rm cm}^{-3}$ in the solar corona;}
\item{$n_e\sim 1\,{\rm cm}^{-3}$ in the ISM.}
\end{itemize}
--- 38 orders of magnitude variation in density!
Thus, one can readily find many astrophysical systems that 
are not only large but also rarefied and collisionless.

%-----------------------------------------------------------------

\subsection{The Fast Reconnection Condition}
\label{subsec-condition}

How can one quantify condition~(\ref{eq-1}) of collisionless reconnection?
It is pretty straight-forward to show (Yamada~et~al. 2006) that
\beq
{{\delta_{\rm SP}}\over{d_i}} \sim 
\biggl({L\over{\lambda_{e,\rm mfp}}}\biggr)^{1/2}\, 
\biggl(\beta_e\,{m_e\over{m_i}}\biggr)^{1/4} \, ,
\label{eq-2}
\eeq
where I have neglected numerical factors of order~1.
Here, $\beta_e$ is the ratio of the plasma pressure inside the layer to 
the pressure of the reconnecting magnetic field component ($B_0^2/8\pi$)
outside the layer; $\lambda_{e,\rm mfp}$ is the classical electron mean 
free path due to Coulomb collisions. Thus, using equation~(\ref{eq-1}),
we see that reconnection is collisionless when
\beq
\lambda_{e,\rm mfp} > L\sqrt{\beta m_e/m_i} \simeq L\beta^{1/2} /40 \, .
\label{eq-3}
\eeq
[The condition suggested by Yamada~et~al. (2006) differs from 
equation~(\ref{eq-3}) by a factor of~2. Since the discussion 
here is very qualitative, I regard this difference as unessential.
We are not going to quibble about factors of~2, are we?]

We can go a little bit further. The mean free path $\lambda_{e,\rm mfp}$ 
can be written as
\beq
\lambda_{e,\rm mfp} \simeq 7\cdot 10^{7}{\rm cm}\, n_{10}^{-1}\, T_7^2 \, ,
\label{eq-lambda-1}
\eeq
where we have taken the Coulomb logarithm equal to~20 and where~$n_{10}$ 
and~$T_7$ are the electron density~$n_e$ and temperature~$T_e$ given in 
units of $10^{10}\, {\rm cm}^{-3}$ and~$10^7$~K, respectively. These 
parameters are to be taken at the center of the reconnection layer.
Combining equations~(\ref{eq-3}) and~(\ref{eq-lambda-1}), the criterion 
for fast collisionless reconnection can now be formulated as a condition 
on the layer's length~$L$ in terms of the central values of~$n_e$ and~$T_e$: 
\beq
L < L_c \equiv 40\, \beta^{-1/2}\, \lambda_{e,\rm mfp} 
\simeq 3\cdot 10^{9}{\rm cm}\ \beta^{-1/2}\, n_{10}^{-1}\, T_7^2 \, .
\label{eq-L_c-1}
\eeq

Now, what about the plasma-$\beta$ parameter?
For definiteness, let us focus on the extreme 
case of a reconnecting configuration with no
guide field. Then the condition of pressure
balance across the layer dictates that
\beq
\beta \equiv {{8\pi n_e k_B (T_e+T_i)}\over{B_0^2}} = 1 \, ,
\label{eq-pressure-balance}
\eeq
where we have neglected the outside thermal pressure.

Furthermore, assuming for simplicity that $T_e=T_i$, 
we can then express the central electron temperature 
in terms of~$B_0$ and~$n_e$ as
\beq
T_e = {{B_0^2/8\pi}\over{2k_B n_e}} \simeq
1.4 \cdot 10^7\, {\rm K} \, B_{1.5}^2 \, n_{10}^{-1} \, ,
\label{eq-T_e}
\eeq
where $B_{1.5}$ is the outside magnetic field $B_0$ expressed in units 
of~30~G. Upon substituting this estimate into equation~(\ref{eq-lambda-1}), 
we get
\beq
\lambda_{e,\rm mfp} \simeq 1.5\cdot 10^8{\rm cm}\, n_{10}^{-3}\, B_{1.5}^4 \, ,
\label{eq-lambda-2}
\eeq
and correspondingly, 
\beq
L_c(n,B_0) \simeq 6\cdot 10^{9}{\rm cm}\, n_{10}^{-3}\, B_{1.5}^4 \,  .
\label{eq-L_c-2}
\eeq

We see that the condition $L< L_c$ is easily satisfied for solar flares, 
for example. Also, this result has interesting implications for the coronal 
heating problem, as I will discuss in the Appendix.

A reservation: this was just a simple example, presented in 
order to illustrate the basic idea. In general, the physics
is more complicated and less certain. In particular, in the 
above example I have assumed the plasma pressure at the center 
of the current layer to be equal to the outside magnetic pressure 
outside (the background gas pressure in the corona outside the 
layer is negligible since the corona is almost force-free), i.e., 
that $\beta\simeq 1$. However, if there is a guide magnetic field, 
then the cross-layer pressure balance is modified; in particular, 
$\beta$ can be much less then~1, determined by thermal transport 
processes along the layer, e.g., the electron thermal conduction 
(radiative losses are small on the timescale of transit through
the layer). Just as important, since collisions are rare, and 
since the collisional electron-ion energy-equilibration rate is 
suppressed due to the large mass ratio, the electron and ion 
temperatures in the layer need not be equal. For example, ions 
may be much hotter than the electrons and may provide the bulk 
of the pressure support against the outside magnetic field. 
The electron temperature in this case would be far below the 
equipartition value (about $10^7$ K). Correspondingly, the electron 
mean-free path would be much lower than that given by equation~(\ref
{eq-lambda-2}).

%----------------------------------------------------------------

\subsection{Fast Reconnection: Caveats and Alternatives}
\label{subsec-caveats}

With all this said, there is still room for caveats and
alternative ideas. I will mention just some of them here:

\begin{itemize}

\item{Recent numerical work by Cassak~et~al. (2005) shows 
an intriguing evidence for {\bf bistable reconnection}: 
once fast Hall reconnection has begun, it is hard to switch 
it back to the slow Sweet--Parker mode, even if the resistivity 
is raised to the level that violates (\ref{eq-1}).}

\item{{\bf Turbulent Reconnection:} Lazarian \& Vishniac (1999)
suggested that fast reconnection may happen in pure resistive MHD, 
although only in 3D, in the presence of externally imposed MHD turbulence
(see also Bhattacharjee \& Hameiri 1986; Strauss~1988; Kim \& Diamond~2001).

We don't know really whether this mechanism works, but certainly it 
is an interesting idea. And a testable one! Numerical tests should 
now be possible. It is worth pursuing and may be a good topic for a 
PhD thesis project!
}

\item{{\bf Bursty, impulsive reconnection:} e.g., Bhattacharjee (2004)}

\item{{\bf 3D reconnection}: e.g., Longcope (1996)}

\item{{\bf Additional Physics:} e.g., reconnection in partially-ionized 
plasmas in the context of molecular clouds (e.g., Zweibel~1989).}

\end{itemize}

%*****************************************************************

\section{Energetically-Important Reconnection}
\label{sec-energy}

%----------------------------------------------------------------

\subsection{Energetically-Important Reconnection: Go where the Energy is!}
\label{subsec-energy}

In addition to the question ``where can it occur?'', one also 
can ask the question: {\it where does fast reconnection matter?} 

Often in Astrophysics, when judging the relative importance 
of various components of the system, one tends to look at 
{\it where the energy is}. From this point of view, one can 
say that reconnection matters when it results in a {\it significant} 
transfer of magnetic energy to the gas, that is, when the transferred 
energy is larger than the internal in the gas before reconnection. 
Since the energy source in reconnection is the magnetic field, this 
means that we expect for a reconnection event to have a significant 
effect on the system only when the plasma is initially magnetically-dominated, 
$\beta\ll 1$, i.e., the when field is force-free.

This condition, of course, goes in the same direction as the previous 
condition that the system be collisionless and therefore of low density. 
But actually the two conditions are not equivalent, as the force-free 
condition involves the magnitude of the magnetic field, for example.

%----------------------------------------------------------------

\subsection{Energetically-Important Reconnection: 
Externally-Generated Force-Free Field and Gravitational Stratification.}

Generically, where can we expect to encounter a situation described above.

The requirement that the plasma is magnetically-dominated 
means that the magnetic field has to be {\it external} to 
the medium in which reconnection is taking place. That is, 
it needs to be produced in, and anchored by, a denser ($\beta>1$) 
plasma that lies somewhere else, but somewhere nearby. Thus, 
reconnection is important only in situations where one has a 
low-density plasma next to a high-density plasma. This requires 
a strong {\it density stratification}, with the density decreasing 
with distance faster than the magnetic field. This can be most 
easily achieved in the presence of a {\it gravitational field}. 
Thus, we logically arrive to the conclusion that we are dealing 
with magnetospheres or coronae of massive objects with dense gas. 
Two main examples of such systems are: (i) stars and (ii) accretion 
disks.

The basic physical picture is that magnetic field is produced in 
these dense plasmas by a dynamo action and then escapes buoyantly 
into the less dense, largely force-free, corona. At some point in 
the corona, the emerging magnetic loops reach the regions where the 
density is so low that the plasma becomes collisionless. When this 
happens, fast energetic reconnection events (i.e., flares) become 
possible.

%****************************************************************

\section{Observable Reconnection}
\label{sec-observable} 

%----------------------------------------------------------------

\subsection{Observable Reconnection: Optically-Thin Plasma}
\label{subsec-opt-thin} 

The third and final of the three conditions outlined in 
\S~\ref{subsec-outline} is that we want to be able to {\it observe} 
reconnection. Unfortunately, in Astrophysics we do not have the benefit 
of in-situ measurements (as opposed to Space Physics, where reconnection 
events are now routinely being studied with dedicated spacecraft flying 
through the Earth's magnetosphere). Instead, we have to rely on whatever 
radiation that makes its way to us without being greatly absorbed or 
scattered. What this means is that we can observe only those reconnection 
events that happen in an optically-thin (or at least not very optically-thick) 
environment. This statement is, of course, wavelength-dependent. If we are to 
observe the reconnection layer itself, the most relevant part of the spectrum 
is that corresponding to the expected temperature at the center of the 
reconnection layer. This usually falls in the X-ray band; for example, 
it can be as high as $10^7-10^8$~K in the solar corona.

Of course, we would like to be able to directly observe the reconnection 
layer itself. Unfortunately, again, this is frequently not possible. For 
example, even in our best-studied example of astrophysical reconnection
--- the solar flare --- direct observations of the reconnection layer 
itself have not been achieved. And this is so despite the fact that we 
do have X-ray satellites watching the Sun all the time! The reason for 
this is that the reconnection layer is so thin that not much radiation 
comes from it to us directly. An alternative then is to try to see the 
effect of reconnection on the nearby denser medium, which has a direct 
magnetic connection to the reconnection region. Two-ribbon solar flares 
present a classic example of this: the observed white-light radiation 
comes from the photospheric footpoints of newly-reconnected field lines. 
Hard X-rays come from the same footpoints and also from the fast shock 
formed at the spot where the reconnection jet slams into the closed 
post-flare loops; soft X-rays come from these post-reconnection loops.

%--------------------------------------------------------

\subsection{Observable Reconnection: Particle Acceleration}
\label{subsec-particle-acceleration}

The above example illustrates the fact that sometimes (e.g., in solar 
flares) a significant fraction of the dissipated magnetic energy is 
carried away not by the photons, which we could have observed on Earth, 
but by {\it energetic particles}. For example, it is now commonly believed 
that as much as a few tens of percent of the energy released in solar flares 
goes to fast, nonthermal electrons and ions (e.g., Emslie~et~al. 2004; 
Kane~et~al. 2005). Most of the rest of the energy is released as thermal
and bulk kinetic energy of the plasma, with radiative losses playing only 
a minor part in the overall energetics of the reconnection layer.
The mechanisms of particle acceleration by a reconnecting layer are still
not well understood. Nevertheless, this observation implies that reconnection 
processes might also play a role in {\it astrophysical particle acceleration},
e.g., in cosmic-ray production. Physically, what is important here is not just
the large amount of magnetically-stored energy that is rapidly released via 
reconnection, but also the fact that this energy is released in a {\it 
low-density, collisionless environment}. As a result, the nonthermal-particle 
acceleration via whatever mechanism can progress without prohibitive losses
due to collisions. This makes the reconnection phenomenon especially 
important for High-Energy Astrophysics.

%---------------------------------------------------------------

Before I move on to the second part of my talk, I would like to 
reiterate, once again, that I am here talking about large-scale 
reconnection events only. This is not to say that reconnection 
does not occur inside the dense plasma itself (e.g., the solar 
convection zone). But note that: \\
1) The plasma below the solar photosphere is dense and hence collisional; 
therefore, no fast {\it large-scale} and individually-distinguishable 
reconnection events, such as flares, can occur there. Reconnection in 
the convection zone requires a development of very small scales at the
bottom of the turbulent cascade, leading to a continuous and diffuse
magnetic dissipation. (Another reason for a relatively slow rate of 
reconnection in a high-density plasma: reconnection rate usually is 
limited by the Alfv\'en speed, and the Alfv\'en speed is much
smaller in the dense medium.) \\
2) The magnetic field does not dominate the dynamics there, it is below
equipartition with turbulent motions, so, energetically, reconnection is
not very important. \\
3) It is below the photosphere, and hence unobservable directly.

%****************************************************************

\section{Reconnection in Astrophysics: current-sheet formation
by field-line opening and the Helmet Streamer configuration}
\label{sec-recn-astro}

\begin{figure} [t]
\centerline{\psfig{file=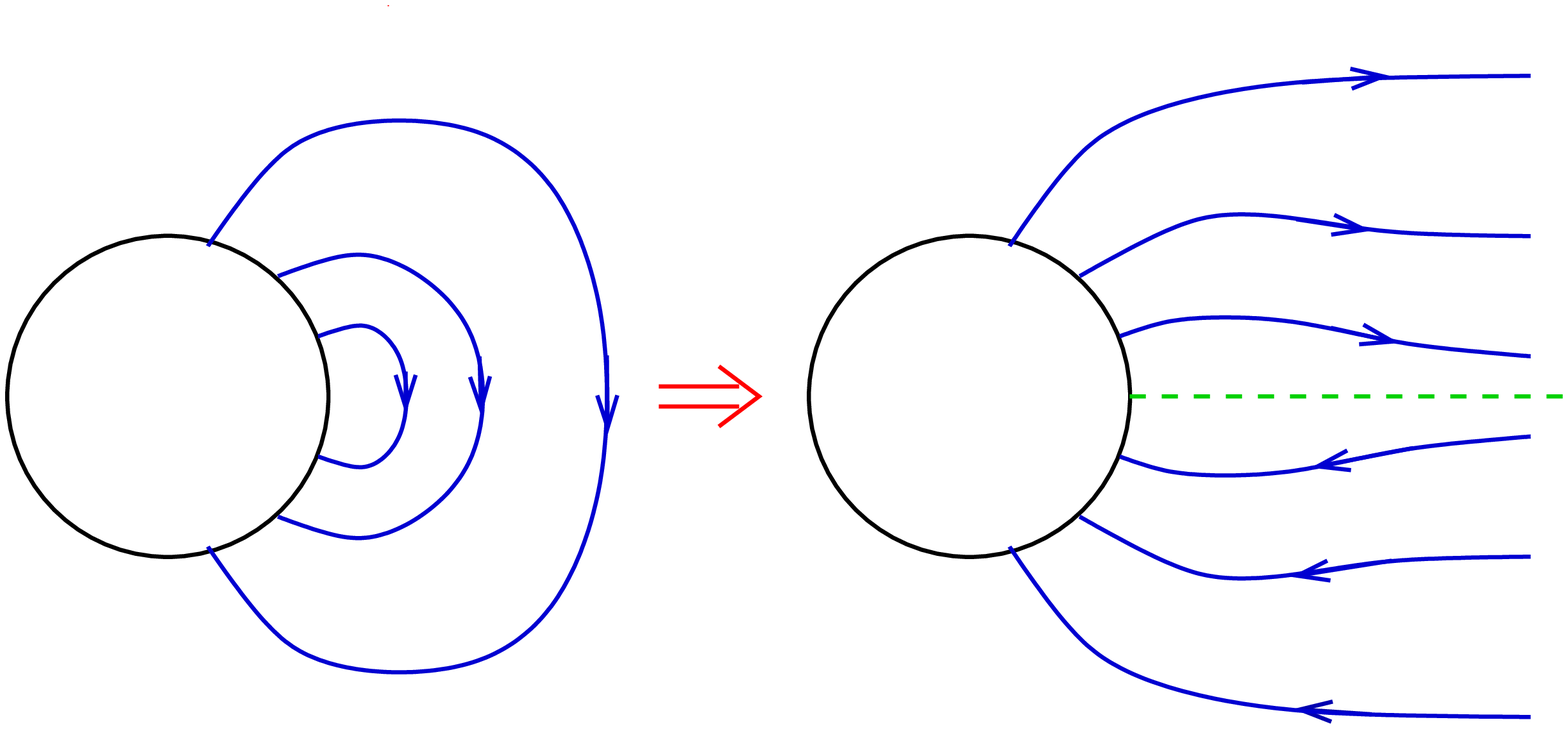,width=4 in}}
\figcaption{Field-line opening from a dipole into a split-monopole
magnetic configuration.
\label{fig-opening}}
\end{figure}

To get reconnection, one first needs to get him/herself a {\it current sheet}.
So, it would be useful to try to understand how a current sheet can develop
in a force-free plasma that is exterior to a denser plasma anchoring the 
magnetic field lines.  

A natural, generic way to get a large-scale current sheet (capable of 
producing a large flare) is the {\it field-line opening}. Indeed, if 
our magnetic field is produced by a dynamo inside the dense conducting 
medium, the field lines that emerge into the low-density corona are
are closed, essentially by design. What this means is that the field
lines are anchored to the dense conductor at both ends, resulting in 
a dipole-like field topology. To get a current sheet, one needs the 
field to become open, that is to go from a dipole to split monopole 
configuration (see Fig.~\ref{fig-opening}).

How does one open field lines? There are several ways to do it,
each one having its own particular application niche:

\begin{itemize}

\item{{\bf non-relativistic force-free case:} 
opening by the sheared motion of the field-line footpoints on the surface
of the dense plasma. Examples: turbulent random walk (solar photosphere); 
differential rotation (accretion disk).}
\item{{\bf relativistic force-free case:} 
opening due to field-line rotation (non-differential!) beyond 
the light cylinder. Examples: pulsar magnetosphere (outer light 
cylinder); the magnetosphere of a disk around a Kerr black hole 
(inner light cylinder). In the first case, the field lines that 
are tied to the rotating pulsar and extend beyond the light cylinder
open out to infinity. This important fact has been known since the 
classical paper by Goldreich \& Julian (1969); however, the actual 
opening process itself was demonstrated in time-dependent numerical 
simulations only recently (see Komissarov~2006; McKinney~2006;  
Spitkovsky~2006).
In the second case, that of a black hole, the field lines that are 
tied to the disk open into the black hole (similar to Koide~2003; 
Beskin~2004; Komissarov~2005).}
\item{{\bf non-relativistic MHD case:} 
field-line opening by a wind. Examples: solar wind; accretion disk wind.} 
\end{itemize}

\begin{figure} [h]
\centerline{\psfig{file=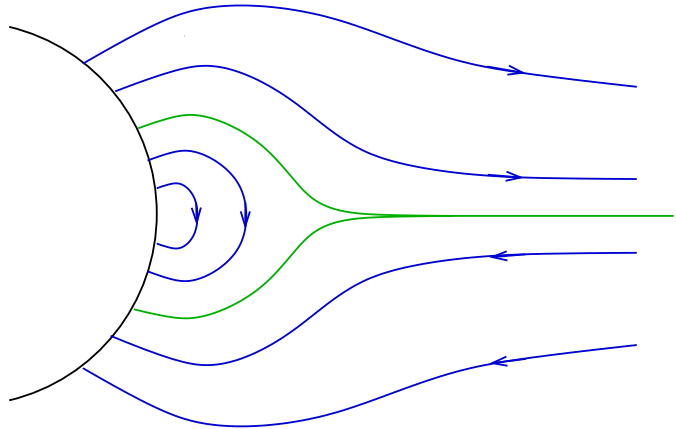,width=3.5 in}}
\figcaption{The Helmet Streamer. The configuration consists of 
two regions of oppositely-directed open field lines and a region 
of closed field lines. The green line represents the magnetic separatrix
current sheet that separates these three regions. 
\label{fig-helmet-streamer}}
\end{figure}

In any case, an important and very generic magnetic structure that one gets 
as a result of a (partial) field-line opening is the {\it Helmet Streamer}
(Fig.~\ref{fig-helmet-streamer}), with magnetic separatrices joining at a 
Y-point (or a {\it cusp-point}; see Uzdensky \& Kulsrud 1997).

%****************************************************************

\section{Examples of Astrophysical Reconnection}
\label{sec-examples}

Examples of astrophysical systems where reconnection is important:

\begin{itemize}

\item{{\bf Stellar corona:} a classic example, very important 
(solar flares); it has been discussed at length, so I am not 
going to talk about it. AT ALL!}

Except, I will mention one special, rather exotic case: 
\item{Relativistic reconnection in the {\bf magnetar magnetosphere}  
(see Fig.~\ref{fig-magnetar}), as a model for {\it SGR flares} 
(Thompson~et~al. 2003; Lyutikov~2003; 2006).}

\begin{figure} [h]
\centerline{\psfig{file=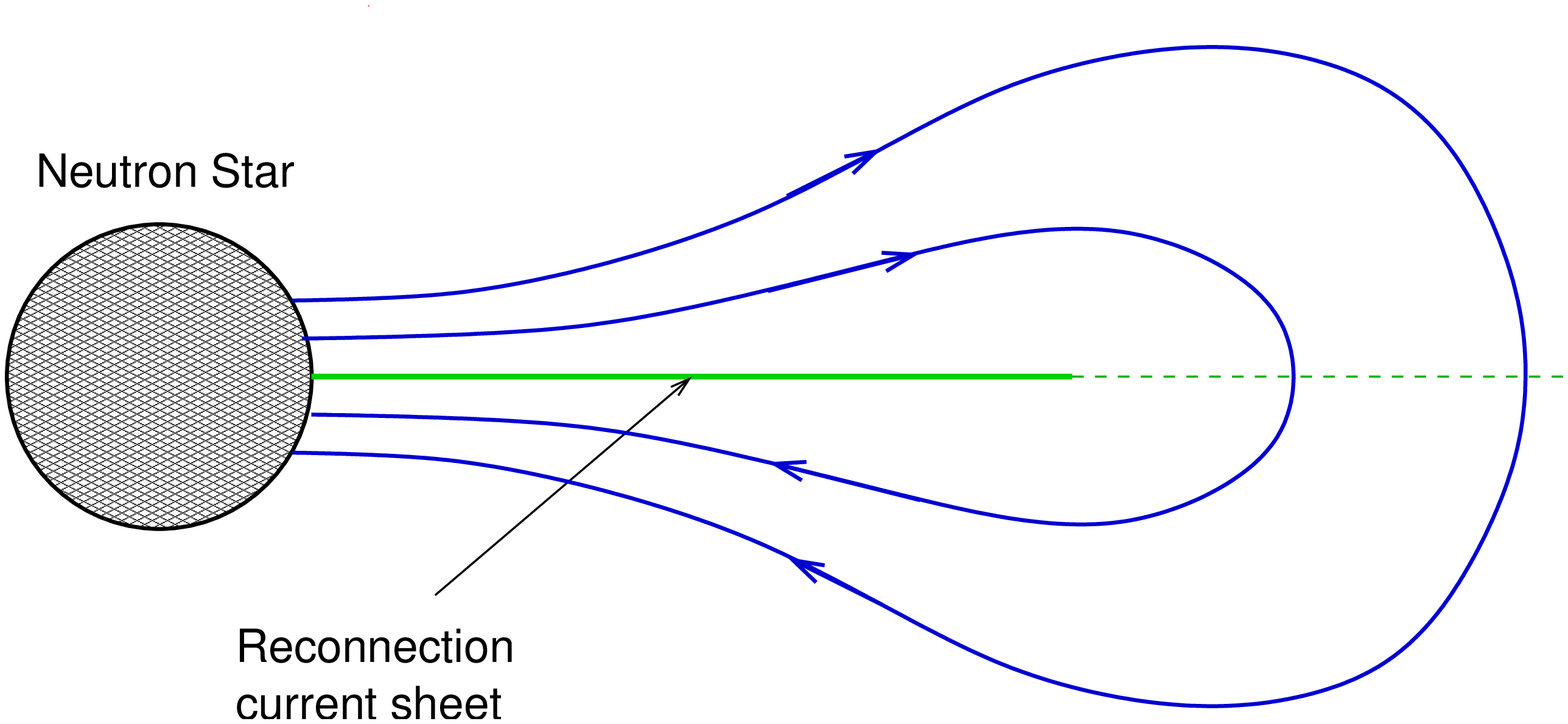,width=4 in}}
\figcaption{Current-sheet formation in the magnetar magnetosphere
(after Lyutikov~2006). 
\label{fig-magnetar}}
\end{figure}

This case represents a manifestation of so-called {\it relativistic magnetic 
reconnection}, in which the Alfv\'en velocity is ultra-relativistic and 
hence special-relativistic effects, such as the Lorentz contraction, play 
a significant role. It has to be noted that relativistic reconnection is 
a relatively new, emerging area of research: only a handful of papers 
devoted to relativistic reconnection have been published to date (e.g., 
Blackman \& Field 1994; Lyutikov \& Uzdensky 2003; Lyubarsky~2005; 
Watanabe \& Yokoyama~2006). Despite the fact that the importance of 
reconnection processes in relativistic plasmas is being recognized 
more and more by the high-energy astrophysics community, this field 
is still not over-crowded. Therefore, in my view, relativistic reconnection 
represents an attractive and promising direction of research. Note that 
the relevant physics can be very rich and perhaps somewhat alien for a 
garden-variety plasma or space/solar physicist. For example, in addition 
to special-relativistic effects, one will have to take into account the 
interaction of plasma with electro-magnetic radiation, e.g., pair 
production/annihilation and very non-trivial radiation transport 
inside the reconnection layer. Nevertheless, I am hopeful that we 
will witness rapid theoretical progress in this exciting and important 
area in the next few years.

%-----------------------------------------------------------

\vskip 20 pt

My main two examples will be magnetospheres and coronae of accretion disks.

\item{{\bf Accretion disk magnetosphere:} star-disk interaction:
current-sheet formation due to differential star--disk rotation 
(\S~\ref{subsec-star-disk}).}

\item{{\bf Accretion disk corona:} reconnection controls magnetic 
scale-height, and hence the vertical extent of the corona 
(\S~\ref{subsec-discorona}).}  

\end{itemize}

Here I distinguish between a large-scale (comparable to the 
global system size) magnetosphere and a corona, formed by 
small-scale (compared with the global system size, i.e., 
the disk radius) magnetic structures.

%-----------------------------------------------------------

\subsection{Reconnection in the Accretion-Disk Magnetosphere: 
Star--disk Magnetic Interaction}
\label{subsec-star-disk}

Let us consider the common star--disk magnetosphere. 
We are interested in the case in which the star has 
a large-scale, dipole-like magnetic field, and we are
interested in the magnetic interaction between this star
and an accretion disk around it (see Uzdensky~2004 for a 
recent review). This fundamental-physics problem is of 
great importance both for accreting Young Stellar Objects 
(YSO), e.g., T Tauri stars, and for accreting compact stars 
in binary systems, e.g., neutron stars (NSs) in X-ray binaries.

During the past 30 years there have been a number of theoretical
studies of such magnetically linked star--disk systems. One of 
the most important early works is that by Ghosh \& Lamb (1978)
in the context of accreting neutron stars. They proposed a steady-state 
model in which the stellar dipole-like magnetic field penetrates 
into the disk over a wide range of radii. This star--disk magnetic
coupling results in the exchange of angular momentum between the star 
and disk, and thus regulates the long-term evolution of the star's 
rotation rate. It also has a significant effect on the accretion flow;
in particular, close enough to the star the magnetic field may become 
so strong that it disrupts the disk completely and channels the accreting 
matter directly onto the polar caps (as is believed to be the case for
X-ray pulsars). Subsequently, K{\"o}nigl (1991) has extended this model 
to YSOs, and this concept of magnetospheric accretion has become a standard 
paradigm in this field as well.

\begin{figure} [h]
\centerline{\psfig{file=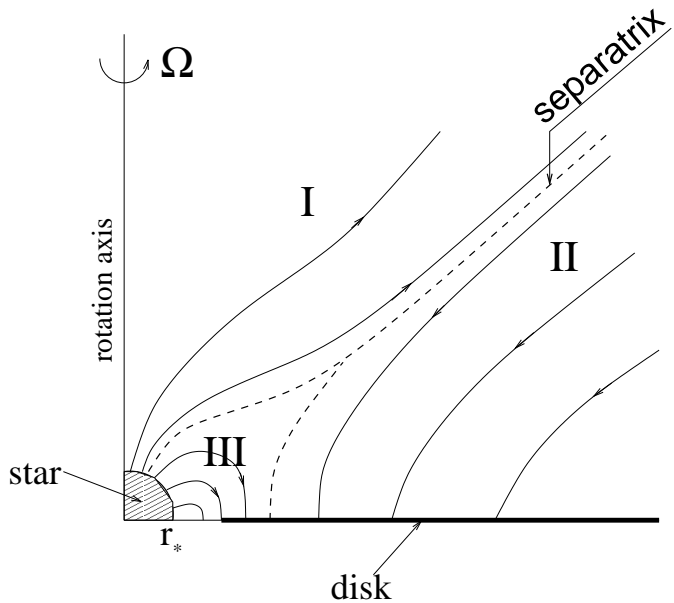,width=3.5 in}}
\figcaption{Current-sheet formation by field-line opening due to 
the differential rotation in the magnetically-linked star--disk
system (after Lovelace et al. 1995). 
\label{fig-lrbk}}
\end{figure}

However, it is easy to see that a steady-state model like this 
requires the disk to have an unrealistically-large resistivity. 
If the resistivity is small (which is more likely), then the 
footpoints of the magnetic field lines can be regarded as being
frozen into the disk on the time-scale of interest. The differential
starr-disk rotation then leads to the twisting of the field lines.
and hence to the generation of the toroidal field in the magnetosphere
above the disk. The pressure of this toroidal field pushes out the 
poloidal field lines, they expand dramatically and effectively open 
up. This line of reasoning has lead Lovelace~et~al. (1995) to propose 
a different steady-state configuration, shown in Figure~\ref{fig-lrbk}. 
In this configuration, the magnetic link between the star and the disk 
has been almost completely severed, with the exception of a small inner 
disk region forced to be in rigid corotation with the star by the strong 
magnetic field.

The process of opening of an axisymmetric magnetosphere subject 
to differential footpoint rotation has been studied extensively
in Solar Physics (e.g., Barnes \& Strurrock 1972; Aly~1984;
Mikic \& Linker~1994). 
There hase been a number of studies of this process in the context 
of accretion disks (van~Ballegooijen~1994; Lynden-Bell \& Boily~1994; 
Lovelace~et~al. 1995; Hayashi~et~al. 1996; Miller \& Stone 1997; 
Goodson~et~al. 1997, 1999; Romanova~et~al. 1998; Uzdensky~et~al. 
2002a,b; Uzdensky~2002a,b; Matt~et~al. 2002; Fendt~2003). 
From the point of view of this talk, the most important feature of 
this opening is the formation of a conical current sheet along the 
separatrix between the oppositely-directed stellar and the disk open 
field lines (see Uzdensky~2002a for more details). This current sheet 
can be clearly seen in Figure~\ref{fig-lrbk}. An important question 
that arises is whether this configuration will persist indefintiely, 
as in the Lovelace~et~al. (1995) steady-state model, or will be subject 
to periodic reconnection events, as was seen in several resistive MHD 
simulations (e.g., Hayashi~et~al. 1996; Romanova~et~al. 1998; Goodson 
et~al. 1999). Another, related question is what would be the amplitude 
and the frequency of the plasmoids that are formed by reconnection and 
ejected along the current sheet (see, e.g., Fendt~2003). These questions 
have important implications for the formation and structure of YSO jets, 
for example. Notice that reconnection processes are at the very heart of 
the difference between the Lovelace et al. (1995) and Goodson~et~al. (1999) 
models. Therefore, it is unlikely that even the most sophisticated resistive 
MHD simulations will be able to settle this issue satisfactory until basic 
physics of reconnection is sufficiently well understood.

%--------------------------------------------------------------------

\subsection{Reconnection in Accretion-Disk Coronae}
\label{subsec-discorona}

Galeev et al. (1979) have proposed a compelling physical picture of how 
a magnetic corona is formed above an accretion disk. The scenario is 
based on the analogy with the solar corona. In both cases, there is a 
dense turbulent medium (the solar convection zone or a turbulent accretion 
disk) whose behavior obeys MHD equations. Magnetic fields are generated by 
the MHD dynamo and amplified to a rough equipartition with the kinetic 
energy density of the turbulence. The medium is in an external gravitational 
field and therefore the plasma density drops off rapidly with height. 
The Parker instability develop and leads to the buoyant rise of magnetic 
flux tubes and their emergence into the lower-density corona above the 
dense medium. Because of the strong gravitational stratification of the 
cold plasma, the energy density in the corona is dominated by the magnetic 
field, which is hence force-free almost everywhere in the corona.
Finally, a portion of the turbulent energy in the disk (or in the convection 
zone) is not dissipated locally, but instead is transported vertically by 
the Poynting flux associated with the rising flux tubes and with the work 
done on the coronal magnetic loops by the footpoint motions. This energy 
builds up as free magnetic energy and is episodically dissipated through 
reconnection events (e.g., flares). Alternatively, the energy may be 
transported vertically by waves that dissipate in the corona via mode 
conversion. In any case, this dissipated energy heats the rarefied coronal 
gas and may also lead to non-thermal particle acceleration.
The basic elements of this picture have been confirmed in 3D MHD 
simulations by Miller \& Stone (2000) and by Machida~et~al. (2000).

\begin{figure} [h]
\centerline{\psfig{file=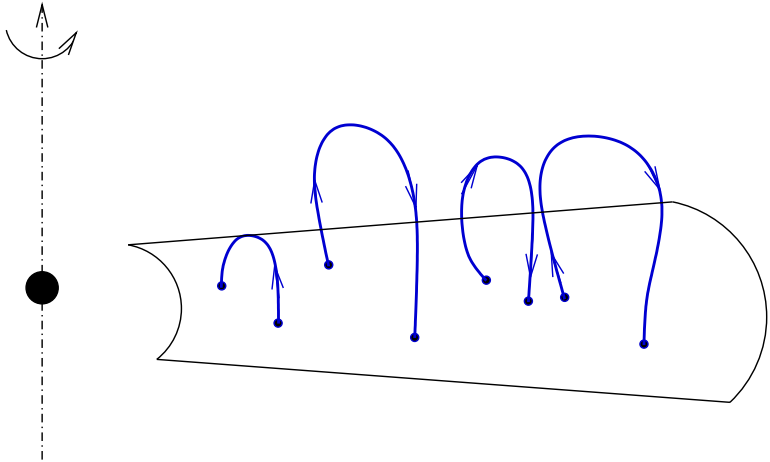,width=4 in}}
\figcaption{Magnetic loops in the corona of a turbulent accretion disk. 
\label{fig-discorona}}
\end{figure}

In order to understand how the coronal magnetic field is structured, 
one needs a statistical description. A promising way to achieve this 
is to represent the coronal field by an ensemble of closed magnetic 
loops of different sizes and to study the distribution function of 
these loops (Tout \& Pringle 1996; Uzdensky \& Goodman 2006).
The evolution of this loop distribution function and its steady-state 
shape are determined by the interplay between the various physical 
processes that govern what happens to the individual magnetic loops 
after they emerge from the surface. In the accretion disk context, 
two of the most important such processes are {\it (i)} the strong shear 
due to the Keplerian differential rotation and {\it (ii)} reconnection 
between loops. The Keplerian shear leads to a rapid stretching of the
loops in the toroidal direction. Correspondingly, the disk motions 
perform work against magnetic forces, which means that magnetic energy 
is pumped into the corona. The coronal flux loops inflate due to the 
increased magnetic pressure and the characteristic magnetic scale-height 
increases. If there were no reconnection, this process would go on until 
the flux tubes grew to a height of order the disk radius~$R$, after which 
the expansion would accelerate and the field would effectively open up. 
Essentially, one would get a dense forest of field lines going up and down.

The main role of reconnection is that it controls the magnetic energy scale 
height of the magnetized corona (e.g., Uzdensky \& Goodman 2006) and acts as 
a mechanism for converting the magnetic energy pumped into the corona into 
the particle energy of hot coronal gas resulting in X-ray emission.
Another important role of reconnection is that it may lead to 
a magnetic ``inverse cascade'' in the corona, i.e., to the production
of a significant population of loops with large {\it radial}%
\footnote
{Loops with large {\it toroidal} footpoint separations are easily
produced by the Keplarian shear alone.}
footpoint separations (see Tout \& Pringle 1996; Uzdensky \& Goodman 2006), 
which may result in an enhanced transport of angular momentum.

%------------------------------------------------------------

\vskip 20 pt

It is my pleasure to thank the Petschek Memorial Symposium organizers, 
Jim Drake and Michael Hesse, and Jim Stone for inviting me to give a talk.

I am very grateful to my teacher Russell Kulsrud for many years of 
his caring mentorship. I learned a lot from Russell over these years!
I also would like to thank Masaaki Yamada and Hantao Ji for many 
useful discussions of physics of reconnection.

This research has been supported by the National Science Foundation 
Grant No.~PHY-0215581 (PFC: Center for Magnetic Self-Organization in 
Laboratory and Astrophysical Plasmas).

%************************************************************

\appendix
\section{Solar Coronal Heating as a Self-Regulated Process}

[The material in this Appendix was not presented at the Meeting.
It has been added only afterwards.]

In this Appendix I discuss the implications of equations~(\ref{eq-L_c-1})
and~(\ref{eq-L_c-2}) for plasma heating in the solar corona.

I propose that coronal heating is a self-regulating process working to keep 
the corona marginally collisionless in the sense of equations~(\ref{eq-3}), 
(\ref{eq-L_c-1}), and~(\ref{eq-L_c-2}). 

As long as the twisting of coronal loops by photospheric
footpoint motions and flux emergence events keep producing 
current sheets in the corona, magnetic dissipation in these
current sheets leads to continuous coronal heating. 
This heating of course is not uniformly distributed but instead 
is localized in both space in time (Rosner~et~al. 1978), as in 
Parker's nano-flare picture (Parker~1988). The overall, integrated 
heating density, i.e., the rate of magnetic dissipation per unit volume, 
depends on the reconnection rate in these sheets, e.g., on whether 
reconnection is fast (e.g., Petschek) or slow (Sweet--Parker).
This is actually a somewhat subtle point. Indeed, in a steady state, 
the energy dissipated in the corona per unit time should be equal to 
the power pumped into the corona magnetically from the solar surface. 
And it is not obvious why and how the energy-pumping rate depends on 
what is happening in the corona. For example, if the corona were enclosed 
in a fixed volume, then the energy dissipation per unit volume would be 
fixed (as in driven MHD turbulence in a box). However, it is important to
recognize that the volume of the corona is not fixed! If, for example,  
reconnection were to suddenly slow down, then more energy would be 
pumped in than the corona could dissipate; then, in order to accommodate 
this additional free magnetic energy, the corona would respond by just 
increasing its scale-height. That is, coronal magnetic structures would 
grow in height until, finally, the total dissipation in the corona became 
equal to the total input from the photosphere. Because of the increased 
volume, the magnetic dissipation per unit volume is decreased.

[Note that there is an additional effect: as coronal magnetic structures
grow in height, the amount of energy pumped into the corona by the footpoint 
motions may go down. This is because the work done by footpoint motions is 
proportional to ${\bf v}_{\rm fp}\cdot {\bf B}_{\rm hor} B_z/4\pi$. As the 
coronal structures grow in height without increasing their lateral size, 
the horizontal field, $B_{\rm hor}$, decreases, whereas the vertical field
component, $B_z$, does not change. Correspondingly, the overall power pumped
from the photosphere into the corona goes down.]

As follows from equation~(\ref{eq-L_c-1}) or equation~(\ref{eq-L_c-2}), 
the regime of reconnection is determined by the global scale~$L$ of the 
reconnection layer and by the basic physical parameters characterizing 
the plasma in the layer (i.e, $n_e$, $T_e$, and~$B_0$). The typical values 
of~$L$ and~$B_0$ are determined by the scale and strength of the magnetic 
structures emerging from the Sun and by the scale of footpoint motions 
(e.g., the meso-granular scale). Therefore, for the purposes of the present
discussion, let us regard~$L$ and~$B_0$ as fixed and ask what determines 
the electron density and temperature in the corona.

Following this line of reasoning, let us invert equation~(\ref{eq-L_c-2})
and view it as the condition for the plasma density. That is, let us 
introduce a scale-dependent critical density, $n_c$, below which the 
reconnection process transitions from the slow collisional Sweet--Parker 
regime to a fast collisionless regime:
\beq
n_c \sim 2\cdot 10^{10}\, {\rm cm}^{-3}\, B_{1.5}^{4/3}\, L_9^{-1/3} \, ,
\label{eq-n_c}
\eeq
where $L_9$ is the global reconnection layer length expressed in 
units of $10^4$~km.

Next, an important link in the chain is the existance of a positive feedback 
between the coronal heating and the density in the corona. This feedback is 
due to the fact that the gas high in the corona comes from evaporation from 
the surface along the field lines that just underwent reconnection. 

Let us consider an example of how this works.

Let us suppose that due to field-line twisting , a reconnecting 
structure is set up in the corona with the current sheet length~$L$ 
and the reconnecting field component~$B_0$. Let us further suppose that, 
initially, the density of the background plasma is higher than~$n_c$, so 
that the reconnection layer is collisional and reconnection proceeds very 
slowly, in the Sweet--Parker regime. That is, there is almost no 
reconnection at all. Coronal heating is then inefficient, the 
surrounding plasma gradually cools and the pressure scale height 
gradually goes down. The gas gradually precipitates. Then the density 
of the plasma entering the layer decreases and at some point becomes 
lower than the critical density. The reconnection process then suddenly 
switches to the collisionless regime. Petschek-like fast reconnection 
ensues, and the rate of magnetic energy dissipation greatly increases. 
A flare commences. Some fraction of the energy released by reconnection 
is transported by the electron conduction along the reconnected field 
lines down to the base, where it is deposited in the dense photospheric 
plasma. This in turn leads to a massive evaporation along the same field 
lines. As a result, the newly-reconnected loops are now populated with 
relatively dense and hot plasma. They cool down only slowly via radiation 
losses, keeping their relatively high density for an appreciable length of 
time. If, during this time, these loops become twisted or somehow get in 
contact with other loops, they are now not likely to reconnect rapidly, 
since their plasma density is above critical. This inhibits further coronal 
heating in the given region. In fact, we can speculate that for any further 
outbursts of coronal activity in the given region to occur, one has to wait 
for the gas in post-reconnective loops to cool down significantly, which 
occurs on a longer, radiative timescale.

Thus we see that, although highly intermittent and inhomogeneous, 
the corona is working to keep itself roughly at about the height-dependent 
critical density given by equation~(\ref{eq-n_c}). Correspondingly, the 
background coronal temperature should be such that results in a density 
scale-height that is just large enough to populate the corona up to the 
critical density level at a given height. In this sense, coronal heating 
regulates itself.

%*****************************************************************

\section*{REFERENCES}
\parindent 0 pt

Aly, J.~J.\ 1984, ApJ, 283, 349

Barnes, C.~W., \& Sturrock, P.~A.\ 1972, ApJ, 174, 659

Beskin, V.~S.\ 2004, J. Korean Phys.Soc., 45, S1711;
preprint (astro-ph/0409076)

Bhattacharjee, A., \& Hameiri, E.\ 1986, Phys. Rev. Lett., 57, 206

Bhattacharjee, A., Ma, Z.~W., \& Wang, X.\ 2001, Phys. Plasmas, 8, 1829

Bhattacharjee, A.\ 2004, ARA\&A, 42, 365 

Birn, J., Drake, J.~F., Shay, M.~A., Rogers, B.~N., Denton, R.~E., 
Hesse, M., Kuznetsova, M., Ma, Z.~W., Bhattacharjee, A., Otto, A., 
\& Pritchett, P.~L.\ 2001, J. Geophys. Res., 106, 3715

Cassak, P., Shay, M., \& Drake, J.\ 2005, Phys. Rev. Lett., 95, 235002

Biskamp, D.\ 1986, Phys. Fluids, 29, 1520.

Biskamp, D., \& Schwarz, E.\ 2001, Phys. Plasmas, 8, 4729. 

Blackman, E.~G., \& Field, G.~B.\ 1994, Phys. Rev. Lett., 72, 494 

Daughton, W., Scudder, J., \& Karimabadi, H.\ 2006, Phys. Plasmas, 13, 072101

Emslie, A.~G., et al.\ 2004, J. Geophys. Res., 109, A10104 

Erkaev, N.~V., Semenov, V.~S., Alexeev, I.~V., \& Biernat, H.~K.\ 2001,
Phys. Plasmas, 8, 4800.

Fendt, C.\ 2003, Ap\&SS, 287, 59

Fujimoto, K.\ 2006, Phys. Plasmas, 13, 072904 

Galeev, A.~A., Rosner, R., \& Vaiana, G.~S.\ 1979, ApJ, 229, 318

Ghosh, P., \& Lamb, F.~K.\ 1978, ApJ, 223, L83

Goodson, A.~P., Winglee, R.~M., \& B{\"o}hm, K.-H.\ 1997, ApJ 489, 199

Goodson, A.~P., B{\"o}hm, K.-H., \& Winglee, R.~M.\ 1999, ApJ 524, 159

Hayashi, M.~R., Shibata, K., \& Matsumoto, R.\ 1996, ApJ 468, L37

Ji, H., Yamada, M., Hsu, S., \& Kulsrud, R.\ 1998, Phys. Rev. Lett., 80, 3256

Kane, S.~R., McTiernan, J.~M., \& Hurley, K.\ 2005, A\&A, 433, 1133

Kim, E., \& Diamond, P.~H.\ 2001, ApJ, 556, 1052

Koide, S.\ 2003, Phys. Rev. D, 67, 104010

Komissarov, S.~S.\ 2005, MNRAS, 359, 801

Komissarov, S.~S.\ 2006, MNRAS, 367, 19

Kulsrud, R.~M.\ 2001, Earth, Planets and Space, 53, 417

Lazarian, A., \& Vishniac, E.~T.\ 1999, ApJ, 517, 700

Longcope, D.~W.\ 1996, Solar Phys., 169, 91

Lovelace, R.~V.~E., Romanova, M.~M., \& Bisnovatyi-Kogan, G.~S.\ 1995, 
MNRAS, 275, 244

Lynden-Bell, D., \& Boily, C.\ 1994, MNRAS, 267, 46
	
Lyubarsky, Y.~E.\ 2005, MNRAS, 358, 113

Lyutikov, M., \& Uzdensky, D.\ 2003, ApJ, 589, 893 

Lyutikov, M.\ 2003, MNRAS, 346, 540

Lyutikov, M.\ 2006, MNRAS, 367, 1594

Malyshkin, L.~M., Linde, T., \& Kulsrud, R.~M.\ 2005, 
Phys. Plasmas, 12, 102902

Matt, S., Goodson, A.~P., Winglee, R.~M., \& B{\"o}hm, K.-H.\ 2002, 
ApJ, 574, 232

McKinney, J.~C.\ 2006, MNRAS, 368, L30

Mikic, Z., \& Linker, J.~A.\ 1994, ApJ, 430, 898

Miller, K.~A., \& Stone, J.~M.\ 1997, ApJ, 489, 890 

Miller, K.~A., \& Stone, J.~M.\ 2000, ApJ, 534, 398

Parker, E.~N.\ 1957, J. Geophys. Res., 62, 509

Parker, E.~N.\ 1988, ApJ, 330, 474

Petschek, H.~E.\ 1964, AAS-NASA Symposium on Solar Flares, 
(National Aeronautics and Space Administration, Washington, 
DC, 1964), NASA SP50, 425.

Romanova, M.~M., Ustyugova, G.~V., Koldoba, A.~V., Chechetkin, V.~M., \&
Lovelace, R.~V.~E.\ 1998, ApJ, 500, 703 

Rosner, R., Tucker, W.~H., \& Vaiana, G.~S.\ 1978, ApJ, 220, 643

Sato, T., \& Hayashi, T.\ 1979, Phys. Fluids, 22, 1189.

Scholer, M.\ 1989, J. Geophys. Res., 94, 8805.

Shay, M.~A., Drake, J.~F., Denton, R.~E., \& Biskamp, D.\ 1998,
J.~Geophys.~Res., 103, 9165

Spitkovsky, A. 2006, accepted to ApJ Letters; preprint (astro-ph/0603147) 

Strauss, H.~R.\ 1988, ApJ, 326, 412

Sweet, P.~A.\ 1958, in IAU Symp.~6, Electromagnetic Phenomena 
in Cosmical Physics, ed.\ B.~Lehnert, (Cambridge: Cambridge Univ. 
Press), 123.

Thompson, C., Lyutikov, M., \& Kulkarni, S.~R.\ 2003, ApJ, 574, 332

Tout, C.~A., \& Pringle, J.~E.\ 1996, MNRAS, 281, 219

Trintchouk, F., Yamada, M., Ji, H., Kulsrud, R.~M. \& Carter, T.~A. \
2003, Phys. Plasmas, 10, 319

Ugai, M., \& Tsuda, T.\ 1977, J. Plasma Phys., 17, 337.

Uzdensky, D.~A., \& Kulsrud, R.~M.\ 1997, Phys. Plasmas, 4, 3960

Uzdensky, D.~A., \& Kulsrud, R.~M.\ 2000, Phys. Plasmas, 7, 4018.

Uzdensky, D.~A., K{\"o}nigl, A., \& Litwin, C.\ 2002a, ApJ, 565, 1191

Uzdensky, D.~A., K{\"o}nigl, A., \& Litwin, C.\ 2002b, ApJ, 565, 1205

Uzdensky, D.~A.\ 2002a, ApJ, 572, 432

Uzdensky, D.~A.\ 2002b, ApJ, 574, 1011

Uzdensky, D.~A.\ 2004, Ap\&SS, 292, 573

Uzdensky, D.~A., \& Goodman, J.\ 2006, in preparation

van~Ballegooijen, A.~A.\ 1994, Space Sci. Rev., 68, 299 

Watanabe, N., \& Yokoyama, T.\ 2006, accepted to ApJ Letters; 
preprint (astro-ph/0607285)

Yamada, M., Ren, Y., Ji, H., Breslau, J., Gerhardt, S., Kulsrud, R., 
\& Kuritsyn, A. 2006, Phys. Plasmas, 13, 052119

Zweibel, E.~G.\ 1989, ApJ, 340, 550

%*****************************************************************

\end{document}